\documentclass[manuscript]{acmart}

\AtBeginDocument{%
  }

\usepackage[utf8]{inputenc}

\usepackage[para,online,flushleft]{threeparttable}
\usepackage[english]{babel}
\usepackage{amsthm}
\usepackage{amsmath}
\usepackage{varwidth}
\usepackage{subcaption}
\usepackage{calc}
\usepackage{blindtext}
\usepackage{multicol}
\usepackage{graphicx}
\usepackage{paralist}
\usepackage[english]{babel}
\usepackage{float}
\usepackage{makecell}
\usepackage{multirow}
\usepackage{amsfonts}
\usepackage{booktabs}
\usepackage{siunitx}
\usepackage{etoolbox}
\usepackage{numprint}
\usepackage[T1]{fontenc}
\usepackage{algorithm}
\usepackage{flushend}
\usepackage[noend]{algpseudocode}
\usepackage{textcomp}
\usepackage{xcolor}
\usepackage{fancyvrb}
\usepackage{paralist}
\usepackage[inline]{enumitem}
\usepackage{booktabs}
\usepackage{color}
\usepackage{colortbl}
\usepackage{flexisym}
\usepackage{array}
\usepackage{xparse}
\usepackage{xspace}
\usepackage[flushleft]{threeparttable}
\usepackage[nobreak]{mdframed}
\usepackage{framed}
\usepackage{pifont}
\usepackage{balance}
\usepackage[most]{tcolorbox}

\newcommand{\eg}{\hbox{\emph{e.g.}}\xspace}

\newcommand{\etal}{\hbox{\emph{et al.}}\xspace}
\newcommand{\wrt}{\hbox{\emph{w.r.t.}}\xspace}

\usepackage{listings}
\usepackage[]{xcolor}

\theoremstyle{definition}
\newtheorem{definition}{Definition}

\newcommand{\toolname}{\textsc{JETO-Mine}\xspace}
\newcommand{\db}{\textsc{JETO-Bench}\xspace}
\newcommand{\openhands}{\textsc{OpenHands}\xspace}
\newcommand{\mycomment}[1]{}

\usepackage[]{xcolor}
\newcommand{\TODO}[1]{\textcolor{red}{#1}\GenericWarning{}{LaTeX Warning: TODO: #1}}\newcommand\todo\TODO

\newcommand{\MODIFIED}[1]{\textcolor{black}{#1}}\newcommand\modified\MODIFIED

\newcommand{\GHilight}{\makebox[0pt][l]{\color{green!20}\rule[-4pt]{1\linewidth}{10pt}}}

\newcommand{\RHilight}{\makebox[0pt][l]{\color{pink}\rule[-4pt]{1\linewidth}{10pt}}}

\algnewcommand{\Inputs}[1]{%
  \Statex \textbf{Inputs:}
  \Statex \hspace*{\algorithmicindent}\parbox[t]{.8\linewidth}{\raggedright #1}
}
\algnewcommand{\Outputs}[1]{%
  \Statex \textbf{Outputs:}
  \Statex \hspace*{\algorithmicindent}\parbox[t]{.8\linewidth}{\raggedright #1}
}
\algnewcommand{\Initialize}[1]{%
  \State \textbf{Initialize:}
  \Statex \hspace*{\algorithmicindent}\parbox[t]{.8\linewidth}{\raggedright #1}
}

\definecolor{javared}{rgb}{0.6,0,0} 
\definecolor{javagreen}{rgb}{0.25,0.5,0.35} 
\definecolor{javapurple}{rgb}{0.5,0,0.35} 
\definecolor{javadocblue}{rgb}{0.25,0.35,0.75} 

\lstdefinestyle{diff}{
    escapechar=\%
}

\usepackage{hyperref}
\addto\extrasenglish{%

}

\usepackage{cleveref}

\setcopyright{none}
\renewcommand\footnotetextcopyrightpermission[1]{} 
\begin{document}

\title{JETO-Bench: A Reproducible Benchmark for Execution Time Improvement Patches in Java}

\author{Khashayar Etemadi}
\email{ketemadi@ethz.ch}
\affiliation{
  \institution{ETH Zurich}
  \country{Switzerland}
}

\author{Zhendong Su}
\email{zhendong.su@inf.ethz.ch}
\affiliation{
  \institution{ETH Zurich}
  \country{Switzerland}
}


\begin{abstract}

Automated fixing of performance issues is gaining attention, but existing benchmarks of execution time improvement patches (ETIPs) target Python, C++, or .NET and are typically fixed datasets that cannot be extended under user-defined configurations. We present \toolname, the first configurable and reusable tool for automatically creating reproducible benchmarks of ETIPs in real-world Java projects. Java is challenging because just-in-time compilation and garbage collection make execution measurements volatile. \toolname employs a three-phase pipeline: static analysis identifies ETIPs from GitHub repositories using user-defined filters and an LLM-based issue classifier, dynamic analysis wraps identified ETIPs in Docker images and performs statistical testing for evidence of execution time improvement, and an evaluation harness supports quantitative assessment of both patch and test generation tools. Using \toolname, we build \db, a benchmark of 660 identified and 91 manually verified executable ETIPs from 174 Java repositories, mined from nearly 1.8 million commits across 11 years. Running \openhands with GPT-5-mini fixes 14.3\% (13/91) of issues, aligning with prior work on other languages. Our results also reveal that open-source Java projects largely lack tests demonstrating execution time improvements, presenting an opportunity for future research in test generation. \toolname and \db are publicly available at \url{https://github.com/khesoem/JETO-Bench}.
\end{abstract}


\maketitle

\section{Introduction}
During the past decade, automated program repair (APR) has made remarkable progress, further accelerated by the advent of agentic coding frameworks~\cite{yang2024swe,wang2025openhands,xia2025demystifying}. Benchmarks such as Defects4J~\cite{just2014defects4j}, QuixBugs~\cite{lin2017quixbugs}, and SWE-Bench~\cite{Jimenez2023SWEbenchCL} have been instrumental in driving these advances, providing standardized tasks and reproducible environments for evaluating APR techniques. However, these benchmarks are almost exclusively focused on \emph{functional} bugs, leaving \emph{non-functional} issues, and performance issues in particular, underrepresented.

Recently, researchers have turned their attention to automated fixing of performance issues~\cite{garg2025rapgen,ren2025peace,yang2025perfcoder,Fan2025SWEEffiRS}. This new trend has led to the creation of benchmarks of execution time improvement patches, such as PIE~\cite{Madaan2023LearningPC}, SWE-Perf~\cite{he2025swe}, GSO~\cite{shetty2025gso}, and PerfBench \cite{garg2025perfbench}. These benchmarks target languages such as Python, C++, and .NET. Despite being one of the most widely used programming languages, Java remains underrepresented in this area.
Note that Java's reliance on just-in-time compilation and garbage collection makes execution time measurements volatile, demanding more rigorous benchmarking methodology than is needed for languages like Python or C++.
The only existing benchmark of execution time improvement patches in real-world Java projects is recently proposed by Yi \etal~\cite{yi2025experimental}, which includes 65 tasks from four repositories. 
This dataset presents a fixed set of patches without containerized environments, making it difficult to reproduce results as project dependencies evolve over time.

In this paper, we present \toolname, the first configurable and reusable tool for creating reproducible benchmarks of real-world execution time improvement patches (ETIPs) in Java. \toolname provides a fully automated pipeline that researchers can run with their own desired filters and configurations to continuously collect new benchmarks. The pipeline consists of three phases. First, a \emph{static analysis} phase crawls GitHub repositories and identifies commits and PRs that address execution time issues, using user-defined filters and an LLM-based issue classifier. Second, a \emph{dynamic analysis} phase wraps the identified ETIPs in Docker images, building and executing the project tests on both original and patched versions to create fully reproducible environments. In this phase, \toolname also runs the existing tests in the project multiple times and conducts statistical testing with user configurable parameters to find objective evidence for improvement of execution time. Third, an \emph{evaluation harness} enables the quantitative assessment of automated patch and test generation tools by running their outputs inside Docker images and performing statistical testing on execution times.

Using \toolname, we build \db, a benchmark of 660 identified ETIPs and 91 manually verified executable ETIPs collected from 174 open-source Java repositories spanning 11 years of development history. \modified{To demonstrate the usefulness of \db, we run \openhands~\cite{wang2025openhands}, a leading open-source coding agent, on the 91 verified executable ETIPs using three advanced LLMs: GPT-5-mini, GLM-5.1, and Mimo-pro-2.5. GPT-5-mini outperforms the other two models by correctly fixing 14.3\% (13/91) of execution time issues, aligning with the results reported by similar studies on other programming languages~\cite{shetty2025gso,garg2025perfbench}.} Furthermore, our analysis reveals that open-source Java projects largely lack tests that quantitatively demonstrate execution time improvements, highlighting a significant gap in current testing practices and a promising opportunity for future research.

In summary, this paper makes the following contributions:
\begin{itemize}
    \item \textbf{\toolname}, the first configurable and reusable tool for automatically creating reproducible benchmarks of execution time improvement patches in Java. \toolname consists of a three-phase pipeline with user-defined configurations and statistical testing.
    \item \textbf{\db}, a benchmark of 660 identified and 91 manually verified executable ETIPs, collected from 174 real-world Java repositories, accompanied by Docker images for full reproducibility.
    \item \modified{\textbf{An empirical evaluation} of \openhands on \db, using GPT-5-mini, GLM-5.1, and Mimo-pro-2.5.} This empirical investigation provides insight into the current capabilities and limitations of advanced coding agents on Java execution time improvement tasks, and reveals the scarcity of execution time improvement tests in open-source Java projects.
\end{itemize}

\section{Execution Time Improvement Patches}
\label{sec:background}

With recent developments in automated bug fixing, researchers are creating new techniques to fix non-functional bugs, such as memory bugs \cite{gao2015safe,ghanavati2020memory}, vulnerabilities \cite{fu2022vulrepair}, code quality deficiencies \cite{etemadi2022sorald}, and execution time issues \cite{selakovic2015poster}. To enable further research in this area, we propose \toolname, the first configurable and reusable tool for creating benchmarks of real-world execution time improvement patches in Java. Execution time improvement patches differ from functional bug fixes in an important way: a correct patch preserves the functionality of the original version while improving its performance \wrt the input under which it is tested. These nuances require precise definitions, which we provide below.

\begin{definition}[Functional Equivalence]
Two versions of a program, original and patched, are \emph{functionally equivalent} if and only if both produce the same output for any expected input.
\end{definition}

\begin{definition}[Execution Time Improvement Patch (ETIP)]
A patch that modifies the original version of a program into the patched version is an \emph{execution time improvement patch (ETIP)} if and only if the two versions are functionally equivalent and the patched version completes its execution significantly faster than the original given a certain expected input. In this case, we consider the original version to have an execution time bug that is fixed by the patch.
\end{definition}

\begin{definition}[ETIP Detector Test]
A test \emph{t} is an \emph{ETIP detector test} if and only if there is an ETIP that modifies an original version P to a patched version Q, \emph{t} runs successfully on both P and Q, and the execution of \emph{t} on Q is significantly faster than on P.
\end{definition}

Note that in our definitions of both ETIP and ETIP detector test, we only consider \emph{significant} improvements of execution time. We emphasize significance in order to dismiss flaky changes due to randomness or execution environment variability. \toolname lets users determine which level of difference is significant by choosing the number of times a test is run, the minimum execution time improvement, and the p-value that entails statistical significance. We explain this user-driven configuration further in \autoref{sec:dynamic-analysis}.

Benchmarking execution time in Java is particularly challenging due to the characteristics of the Java Virtual Machine (JVM). The JVM's just-in-time (JIT) compiler progressively optimizes frequently executed code paths at runtime, meaning that the first several executions of a method can be significantly slower than subsequent ones \cite{georges2007statistically}. Garbage collection (GC) introduces non-deterministic pauses that can distort timing measurements, especially for short running tests \cite{blackburn2006dacapo}. Class loading and bytecode verification also add overhead to early executions. These effects make measurements unreliable for detecting genuine execution time improvements. In contrast with benchmarks targeting Python or C++, where execution time is more directly determined by the source code, a Java ETIP benchmark must explicitly account for JVM warm-up, GC interference, and JIT compilation variability. These JVM-specific concerns motivate several design decisions in \toolname, which we describe in \autoref{sec:dynamic-analysis}.

\toolname identifies ETIPs in open-source Java projects, creates Docker images that enable reproducible execution of these ETIPs, collects test execution time to help quantitative comparison, and provides tools for evaluating generated patches and tests. This framework facilitates future research in the area of execution time improvement in Java programs.

\section{System Design}
\label{sec:system-design}

\subsection{Overview}
\label{sec:overview}

\begin{figure*}
\begin{center}
\includegraphics[width=\textwidth]{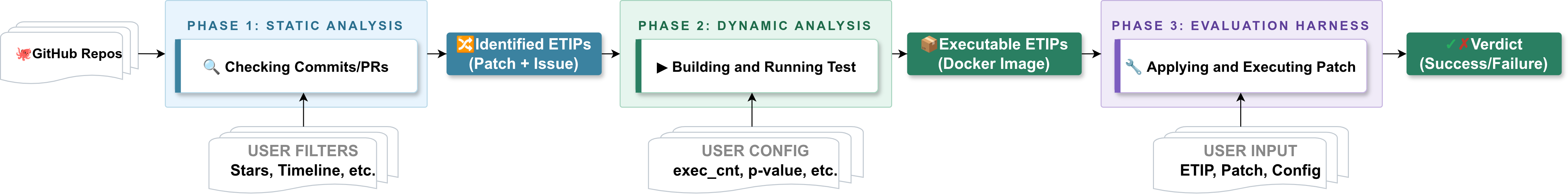}
\caption{An overview of \toolname workflow.}
\label{fig:jpit}
\end{center}
\end{figure*}

\autoref{fig:jpit} illustrates an overview of how \toolname works. The workflow consists of three main phases: 1) Static analysis: \toolname goes over GitHub commits and PRs and identifies ETIPs based on the filters provided by the user, 2) Dynamic analysis: \toolname wraps the identified ETIPs in docker images and creates a fully reproducible environment for their execution, and 3) Evaluation harness: using the reproducible environments generated in the second phase, \toolname provides the tools for the quantitative evaluation of generated patches and tests. In the remainder of this section, we explain each phase in detail.

\subsection{Static Analysis}
\label{sec:static-analysis}

\begin{figure*}
\centering
\begin{subfigure}[b]{0.5\textwidth}
  \centering
  \includegraphics[width=\linewidth,height=7cm,keepaspectratio]{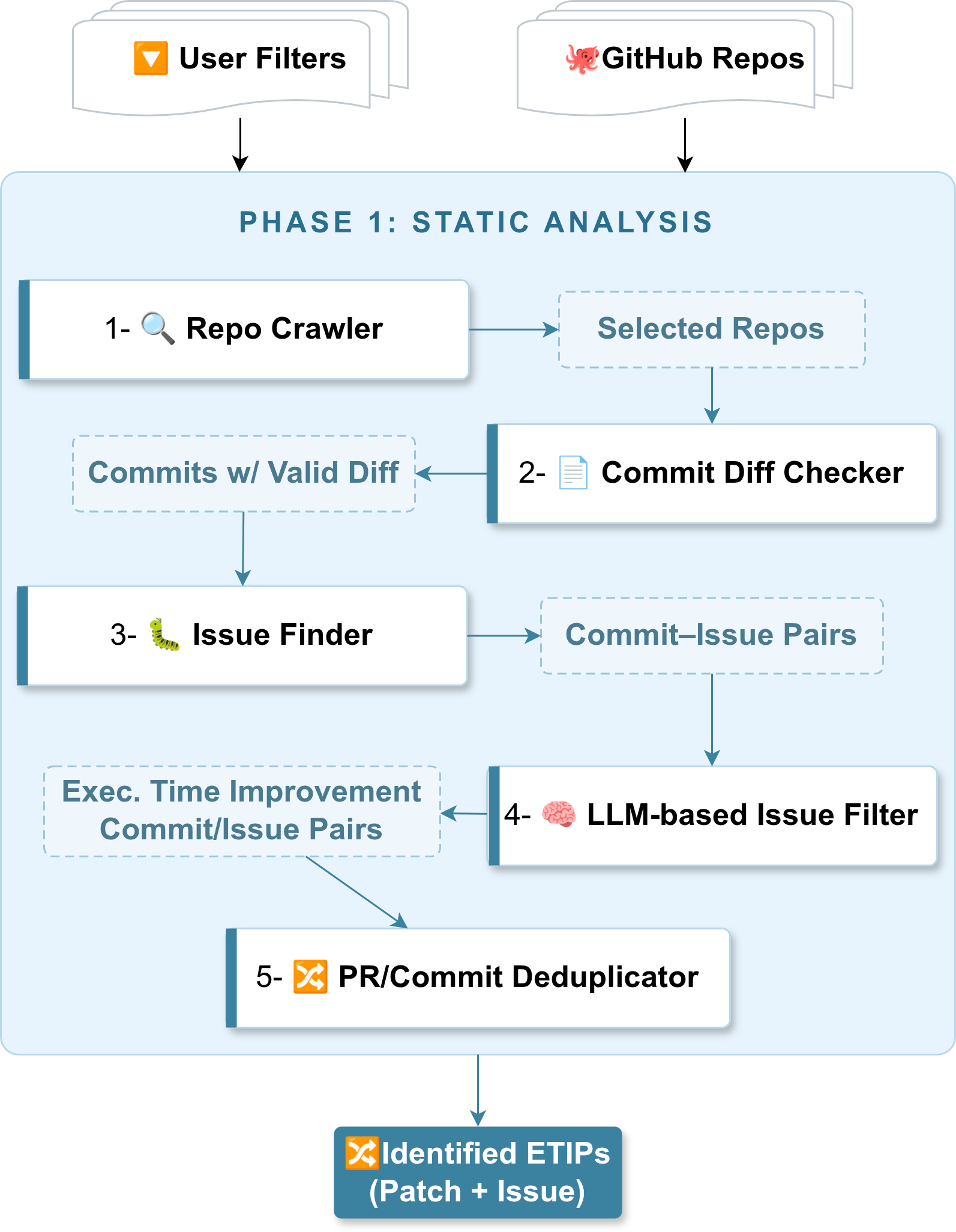}
  \caption{The static analysis phase workflow.}
  \label{fig:jpit-static-analyzer}
\end{subfigure}%
\hfill
\begin{subfigure}[b]{0.5\textwidth}
  \centering
  \includegraphics[width=\linewidth,height=7cm,keepaspectratio]{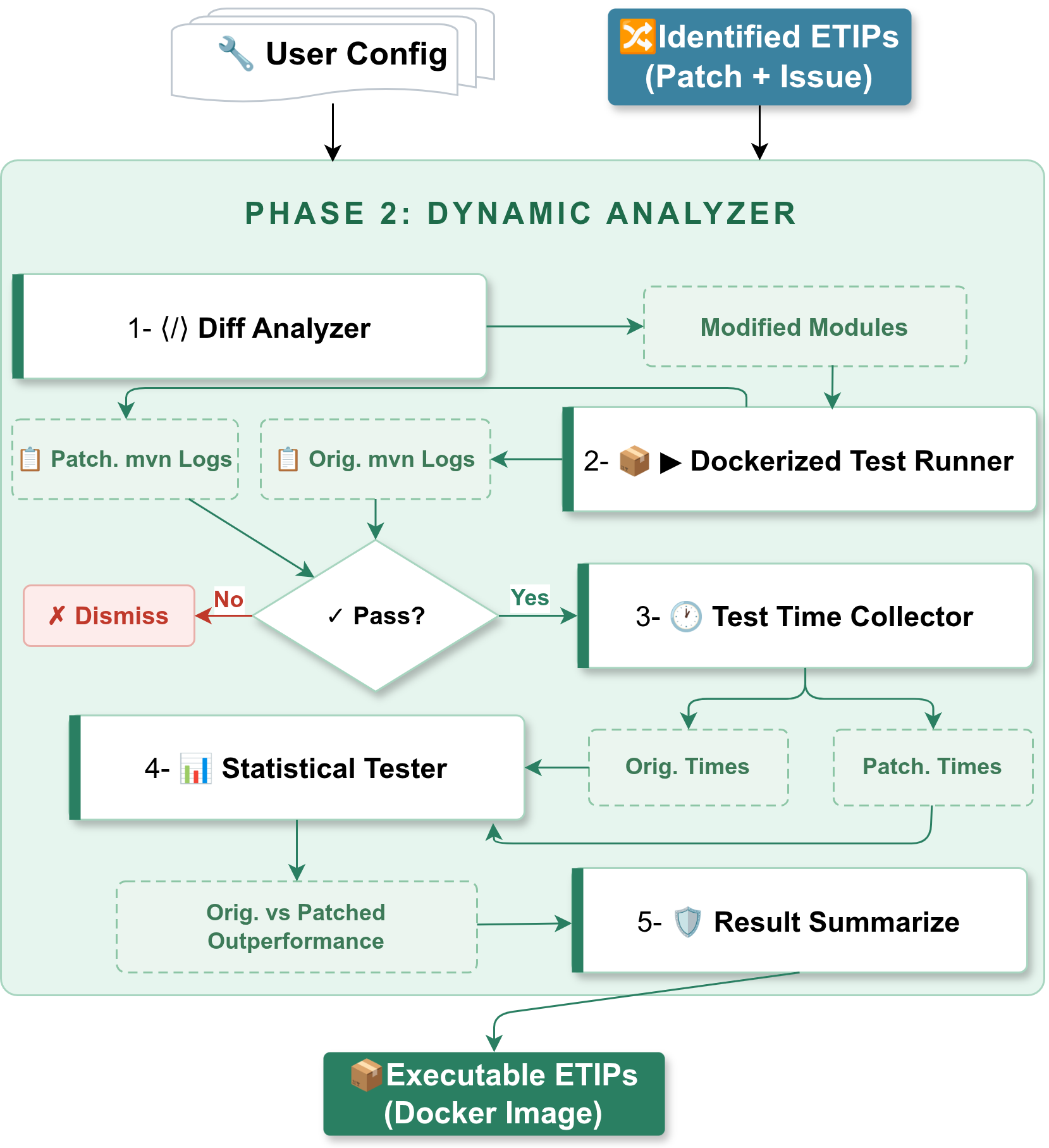}
  \caption{The dynamic analysis phase workflow.}
  \label{fig:jpit-dynamic-analyzer}
\end{subfigure}
\caption{The static and dynamic analysis phases of \toolname.}
\label{fig:jpit-analyzers}
\end{figure*}

\autoref{fig:jpit-static-analyzer} shows how the static analysis phase of \toolname works. The inputs are the user-defined filters that determine which patches should be collected and a GitHub token that enables access to public repositories. After going through patches in the form of both commits and PRs, \toolname identifies and outputs ETIPs and their associated issues. This output provides the necessary data for a performance improvement patch generation task: the original version of the project, the issue that describes the performance problem, and the ground-truth patched version that resolves it.

Static analysis starts by crawling repositories (step 1) that meet four criteria: 1) it is a Java project that uses Java 8 or newer and builds with \texttt{mvnw}, 2) the repository has been updated as recently as the user specified date, 3) it has at least the number of stars determined by the user, and 4) the project is not archived. These criteria ensure that the project is active and popular according to user definition, and uses a build tool that enables reproducibility.


\begin{figure}[t]
\centering
\begin{tcolorbox}[
  colback=gray!5,
  colframe=black!55,
  boxrule=0.4pt,
  arc=1pt,
  left=6pt, right=6pt, top=6pt, bottom=6pt,
  fontupper=\small,
]
The following is an issue in the \textit{$\langle$repo\_name$\rangle$} repository:

\medskip
\#\#\#Issue Title\#\#\#\textit{$\langle$issue\_title$\rangle$}\#\#\#Issue Title End\#\#\#

\#\#\#Issue Body\#\#\#\textit{$\langle$issue\_body$\rangle$}\#\#\#Issue Body End\#\#\#

\medskip
The following is the commit message that fixes this issue:

\medskip
\#\#\#Commit Message\#\#\#\textit{$\langle$commit\_message$\rangle$}\#\#\#Commit Message End\#\#\#

\medskip
Is this issue focused on improving execution time? Answer by only one word: `yes' or `no' (without any other text or punctuation). If you do not have enough information to decide, say `no'.
\end{tcolorbox}
\caption{Prompt used by \toolname{} to classify whether a GitHub issue targets an execution time improvement.}
\label{fig:et-classification-prompt}
\end{figure}

After the desired repositories are selected, \toolname checks the commit diffs on the default branch over the timeline specified by the user (step 2). A commit diff is considered valid if it 1) only changes \texttt{.java} files, 2) does not change test files, and 3) modifies fewer files than the user specified limit. \modified{We consider these criteria as we aim to facilitate the evaluation of patch generation capabilities of automated tools particularly on source code.} The commits that meet these requirements are then checked to find their associated issues (step 3). \toolname then prompts an LLM with the issue title, description, and commit message to determine if the issue is primarily focused on improving execution time (step 4). Only valid commits linked to execution time improvement issues are kept.

\modified{\autoref{fig:et-classification-prompt} shows the prompt used to classify the issues linked to valid commits. The prompt asks the model to label the issue as not focused on improving execution time, if the given information is not enough to decide. This instruction is included to minimize the number of false positives. We prioritize minimizing false positives over false negatives, as we do not want to have too many issues that need to be manually detected and excluded later. We also provide the commit message in the prompt as our preliminary experiments show that affordable models are usually too conservative; they tend to state that not enough information is provided in issue title and issue body. We include the message of the fixing commit to help the model make an informed decision about the issue.}

In the fifth and final step, \toolname cleans up the filtered commit-issue pairs. First, for commits that are part of a PR, we replace the commit with the corresponding PR to capture the complete change intended by the developer. This ensures that the considered code change fully addresses the execution time issue. We also verify that the PR's changed files pass all the considered criteria. Second, we deduplicate the set of all commits and PRs based on their code diff, as there can be multiple commits on the default branch with the same diff depending on the PR merge strategy used in the project.

The result of static analysis is the set of identified ETIPs, each accompanied by an issue describing the execution time problem resolved by the patch. Each ETIP provides three essential pieces of information: an original version of the program with an execution time defect, a GitHub issue describing that defect, and a ground-truth patch showing how the issue can be resolved. Traditionally, such data has been successfully used for the evaluation of patch generation tools~\cite{tufano2019empirical}, which confirms the usability of the identified ETIPs.

Identified ETIPs can be used for patch generation evaluation by checking if a generated patch exactly matches the ground truth~\cite{tufano2019empirical,samant2025syntax}. However, a generated patch can still correctly improve execution time without exactly matching the ground truth, \eg by applying the same algorithmic optimization with minor differences in boundary checks or operation ordering. To provide more fine-grained assessment beyond syntactic similarity, \toolname makes ETIPs executable in a fully reproducible manner. By executing a generated patch, we can determine if it compiles, passes tests, or improves test execution times. Such feedback enables more precise evaluation and can also serve as a signal to iteratively improve the generated patch~\cite{ye2022neural}. To guarantee executability, the identified ETIPs are passed to the dynamic analysis phase of \toolname.

\subsection{Dynamic Analysis}
\label{sec:dynamic-analysis}

The dynamic analysis phase of \toolname, as shown in \autoref{fig:jpit-dynamic-analyzer}, takes the identified ETIPs and user configurations as input and outputs executable ETIPs in the form of Docker images. The resulting images contain the environment needed for building the project and running relevant tests, including all required dependencies, enabling full reproducibility for running generated patches or tests.

The user can configure three parameters that determine whether a patch shows significant execution time improvement: the number of executions $num\_exec$, the minimum improvement percentage $min\_impr$, and the target p-value $p\_val$. These parameters let the user control the statistical rigor of the results.

The dynamic analysis phase of \toolname is designed to address the JVM-specific challenges described in \autoref{sec:background}. To mitigate the effect of JIT compilation warm-up, \toolname executes the test suite once as a dedicated warm-up round before beginning timed executions. The results of the warm-up round are discarded. To reduce the impact of garbage collection pauses and other sources of timing variation, \toolname runs each test suite $num\_exec$ times (30 by default) and applies statistical testing. Moreover, \toolname runs original and patched versions inside the same docker image with identical JVM versions and configurations, ensuring that JIT compilation behavior, GC settings, and class loading conditions are the same for both versions. These measures are essential for producing reliable execution time comparisons on the JVM.

Given the inputs, the dynamic analysis starts by analyzing the code diff between the original and patched versions to identify the modified files and modules (step 1). Next, \toolname creates a Docker image with the correct Java version and builds and runs the JUnit tests of the modified modules on both versions (step 2). The tests are executed using Maven Wrapper multiple times according to the user given configuration. The build and test logs are saved in the image for later investigation and analyzed to ensure that across all executions of both versions: 1) the project successfully builds, 2) the modified modules contain tests, and 3) all tests pass. If any of these requirements is not met, the ETIP is excluded from the final set of executable ETIPs. This ensures that the tests can later be used to assess the functional correctness of generated patches.

After confirming that both versions pass all tests, \toolname extracts test execution times from the logs (step 3). The extracted data consists of times $T_{ijv}$, where $1 \leq i \leq n$ refers to the $i$th test class in the modified modules, $1 \leq j \leq num\_exec$ indicates the execution round, and $v \in \{original, patched\}$ indicates the version.

At step 4, \toolname performs statistical testing on the collected times. For each test class $1 \leq i_{0} \leq n$, \toolname conducts a paired one-sided binomial test between the lists of $T_{i_{0},j,original}$ and $T_{i_{0},j,patched}$ for $1 \leq j \leq num\_exec$. If the test shows that one version is $min\_impr$ percent faster than the other with a p-value below $p\_val$, \toolname considers the $i_{0}$th test class an ETIP detector test. Note that for a given ETIP, some tests may favor the original while others favor the patched version.

Finally, \toolname summarizes the execution results in a report (step 5). For each ETIP, the report indicates whether it successfully builds and passes tests, which test classes detect a significant execution time difference, and whether there is a significant difference in overall execution time. \toolname considers an ETIP executable and creates a Docker image for it if both versions build and pass tests in all $num\_exec$ rounds. Even if no existing test detects an execution time improvement, the Docker-based executability still provides valuable feedback on build and test success that patch generation tools can leverage.

\subsection{Evaluation Harness}
\label{sec:evaluation-harness}

The third and final phase of \toolname provides an evaluation harness. This harness takes the ID of an executable ETIP, a generated patch, and an execution configuration as input, and evaluates whether any test class in the modified modules detects a significant execution time improvement by the generated patch.

The evaluation harness works in five steps. First, the Docker image corresponding to the selected ETIP is pulled if not already present and a container is created. Second, the generated patch is applied to the original version and saved in a separate directory inside the container. Third, the tests are executed on both original and patched versions as many times as specified in the configuration. Fourth, the logs are analyzed to verify that both versions pass all tests in every execution round and to extract test execution times. Finally, a statistical test is performed to determine if any test class detects a statistically significant difference between the execution times of the two versions. The logs and statistical test results are presented to the user as the final report.

In addition to patch evaluation, \toolname also supports the evaluation of test generation tools. For this purpose, \toolname takes a generated test class and runs it on the original and ground-truth patched versions, then conducts statistical testing to determine if the generated test is an ETIP detector test. This enables research in a rarely studied area: the generation of tests that detect execution time improvement.

\subsection{Implementation}
\label{sec:implementation}

We implement \toolname in Python. In its default configuration, the static analysis phase targets projects with at least 20 stars and commits that change at most 20 files since 2015, and uses \texttt{gpt-5.1-codex-mini} for the LLM-based issue filter. In the dynamic analysis phase, the tests are executed once for warm-up and 30 times for collecting execution times. Following previous work~\cite{he2025swe}, \toolname considers an improvement of 5\% with a p-value of 10\% significant. \toolname, \db, and all data relevant to this paper are publicly available~\cite{repo}.

\section{Experimental Methodology}
\label{sec:methodology}

\subsection{Building \db}
\label{sec:building-benchmark}

\begin{table}[ht]
\centering
\begin{tabular}{lrrrrr}
\hline
Metric & Min & Q1 & Median & Q3 & Max \\
\hline
Commits & 1 & 11 & 58 & 288 & 55,824 \\
Stars & 20 & 29 & 53 & 139 & 93,448 \\
\hline
\end{tabular}
\caption{The number of commits and stars across the considered repositories.}
\label{tab:commit_star_summary}
\end{table}

We start our experiments by running \toolname to build \db, the first benchmark of execution time improvement patches in diverse real-world Java projects.

We build \db by running \toolname with its default configuration (see \autoref{sec:implementation}), collecting commits and PRs until 2025-11-28. \toolname checks 3,686 repositories and 1,769,958 commits spanning 11 years of open-source development history. This shows the large scale of \db and our study, which provides unprecedented insight into execution time improvement in open-source Java projects.

\autoref{tab:commit_star_summary} reports the distribution of commits and stars across the considered repositories. The number of commits per repository ranges from 1 to 55,824 with a median of 58, while stars range from 20 to 93,448 with a median of 53, showing notable diversity in both activity and popularity. The result of running \toolname on this large and diverse set of repositories and commits is \db.

\db includes a set of ETIPs identified by the static analysis phase and a set of executable ETIPs in the form of Docker images built by the dynamic analysis phase. We conduct experiments to study the characteristics of \db, the effectiveness of advanced coding agents on \db, and the performance of \toolname in building Java ETIP benchmarks.

\subsection{Research Questions}
\label{sec:rqs}

\newcommand\rqone{What are the characteristics of the ETIPs collected by \toolname in terms of diversity, scope, and test coverage?}
\newcommand\rqtwo{How effectively can advanced coding agents fix execution time issues in \db?}
\newcommand\rqthree{How precise and practical is \toolname for building Java ETIP benchmarks?}

Our experiments study the following research questions:
\begin{itemize}
    \item \textbf{RQ1} (\db characteristics): \rqone \ We assess the characteristics of the identified and executable ETIPs in \db to show the diversity of real-world execution time improvements collected by our approach.
    \item \textbf{RQ2} (using \db in practice): \rqtwo \ We run \openhands, a leading open-source coding agent, on the executable ETIPs in \db to evaluate the current state of automated ETIP generation. We evaluate \openhands performance employing three advanced LLMs: GPT-5-mini, GLM-5.1, and Mimo-pro-2.5.
    \item \textbf{RQ3} (\toolname performance): \rqthree \ We investigate the precision and accuracy of \toolname and measure the cost of benchmark building in terms of time, space, and financial expenses.
\end{itemize}

\subsection{Protocol for RQ1 (\db characteristics)}
\label{sec:rq1-protocol}

To answer \textbf{RQ1}, we count the number of identified ETIPs, executable ETIPs, and the repositories from which they are collected. These numbers indicate the overall applicability of \toolname. We then characterize the ETIPs along four dimensions: the number of files changed, the number of modules modified, the year of creation, and the popularity of the corresponding repository. A diverse set of ETIPs enables the evaluation of various aspects of patch generation tools.

We also examine ETIP detector tests using two statistical testing configurations: 1) the default configuration, following previous work~\cite{he2025swe}, with $p\_val=0.1$ and $min\_impr=0.05$, and 2) a conservative configuration with $p\_val=0.05$ and $min\_impr=0.1$. For each configuration, we check how many executable ETIPs have a test that detects a significant execution time improvement. Cases where no such test exists indicate that the open-source project lacks tests demonstrating execution time improvements, highlighting an opportunity for future research in test generation.

\subsection{Protocol for RQ2 (using \db in practice)}
\label{sec:rq2-protocol}

To answer \textbf{RQ2}, we use \openhands to generate patches for the execution time issues related to executable ETIPs. \openhands provides isolation via Docker sandboxing; we leverage this by letting \openhands use the Docker images created by \toolname as its sandbox environment with all the dependencies and settings needed to run the project. \modified{Inspired by our preliminary experiments, we run \openhands with three models \texttt{gpt-5-mini}, \texttt{glm-5.1}, and \texttt{mimo-pro-mini}. These models provide strong effectiveness at reasonable cost. For each executable ETIP, we give \openhands 100 iterations to fix the issue.}

\modified{To ensure a high-quality evaluation set, we only use executable ETIPs that our manual inspection confirms that they are focused on improving execution time. The details of this inspection are presented in the answer to RQ3 (see \autoref{sec:rq3-protocol} and \autoref{sec:rq3-res}). This yields a manually verified set of executable ETIPs that we recommend to be used for future research as well.}

\modified{After running \openhands, we manually classify generated patches into four categories:} 1) \emph{exact match:} the AST is identical to the ground truth, 2) \emph{semantically equivalent:} functionally equivalent to the ground truth and applying similar optimization but syntactically different, 3) \emph{correct location:} modifies the right code location but differs semantically from the ground truth, and 4) \emph{wrong location:} modifies an incorrect location. The first two categories constitute correct fixes. Although correct location patches fall short of correctness, they demonstrate accurate fault localization, a key step in patch generation \cite{zhang2023survey}. We also track multi-file and multi-module patches in each category to assess performance on issues that span beyond a single file.

We use the evaluation harness to build and run tests on all generated patches. The harness automatically identifies patches that have incorrect format, do not compile, or fail tests, allowing them to be dismissed without manual assessment. We verify that all correct and semantically equivalent patches pass the tests, which reaffirms the soundness of the harness. With this experiment, we check that the evaluation harness of \toolname successfully employs the executable ETIPs in \db to provide execution-based feedback on generated patches.

Note that the goal of this experiment is not to build a state-of-the-art ETIP generation tool. The goal is to investigate whether \db and the evaluation harness can be successfully used to evaluate patch generation tools, and to gain insight into the capabilities of an advanced off-the-shelf agent on Java execution time issues.

\subsection{Protocol for RQ3 (\toolname performance)}
\label{sec:rq3-protocol}

\modified{To answer \textbf{RQ3}, we manually analyze the issues corresponding to all identified ETIPs in \db as well as 100 randomly selected issues that are passed to the LLM-based Issue Filter of \toolname (see \autoref{fig:jpit-static-analyzer}) to measure the precision and accuracy of our tool. To have a solid and clear basis for our analysis, we consider an issue focused on execution time improvement if it meets three requirements: 1) the code modification needed for resolving the issue should inherently lead to an improvement in execution time, 2) the resulting execution time improvement should not be dedicated to improving test execution time, it should be a general improvement to the code, and 3) the code modification needed for resolving the issue should not change the program functionality in terms of input/output.}

\modified{To measure the precision of \toolname, we analyze all issues linked to the identified ETIPs in three steps. First, one of the authors manually checks whether the issue is focused on execution time improvement. Second, we prompt GPT-5.5, one of the most advanced existing LLMs as of July, 2026, and provide it with the issue title and description as well as the three aforementioned requirements and ask the model if the issue is focused on execution time improvement. If the manual label and the label by GPT-5.5 are the same, we take it as the final ground-truth label of the issue; otherwise, the issue goes to the last step. In the third and last step, if the manual label and the label by GPT-5.5 are not aligned, we ask another expert to manually label the issue as a tie breaker. All participating experts in this experiment have more than five years of programming experience. The ratio of issues that are labeled as focused on execution time improvement determine the precision of \toolname.}

\modified{To assess overall accuracy, we randomly sample 100 issues passed to the LLM-based Issue Filter of \toolname (see \autoref{fig:jpit-static-analyzer}). We follow the same three step process as used for precision measurement to assign ground-truth labels to the 100 sampled issues. Then, we check whether the label assigned by \toolname aligns with the ground-truth label. The ratio of the issues for which the labels align is the measured accuracy of \toolname.}

We also measure the time needed for identifying ETIPs, the cost of LLM-based issue filtering, the time the dynamic analysis phase takes to build Docker images, and the size of the resulting images. These numbers indicate the feasibility of using \toolname on a large corpus of open-source commits.

\subsection{Experiment Resources}
\label{sec:resources}

We run all experiments on a machine with one AMD EPYC 7742 processor (64 cores, up to 2.25\,GHz) and 256\,GB DDR4 RAM (3200\,MT/s). In the dynamic analysis phase, we dedicate 32 CPU cores and 80\,GB RAM for building each Docker image.

\section{Experimental Results}

\subsection{Results for RQ1 (\db characteristics)}
\label{sec:rq1-results}

\db contains 660 identified ETIPs and 102 executable ETIPs\footnote{Of the 102 executable ETIPs, 91 pass our manual verification criteria (see \autoref{sec:rq3-protocol} and \autoref{sec:rq3-res}).}. Compared to similar benchmarks for other programming languages \cite{garg2025perfbench,he2025swe}, 102 executable patches provide sufficient data for evaluating advanced patch generation techniques. Additionally, 21 ETIPs build successfully but are excluded from the executable set because their modified modules contain no tests, showing that the dynamic analysis effectively filters out patches lacking functional correctness validation.

\begin{figure}
\begin{center}
\includegraphics[width=0.7\columnwidth]{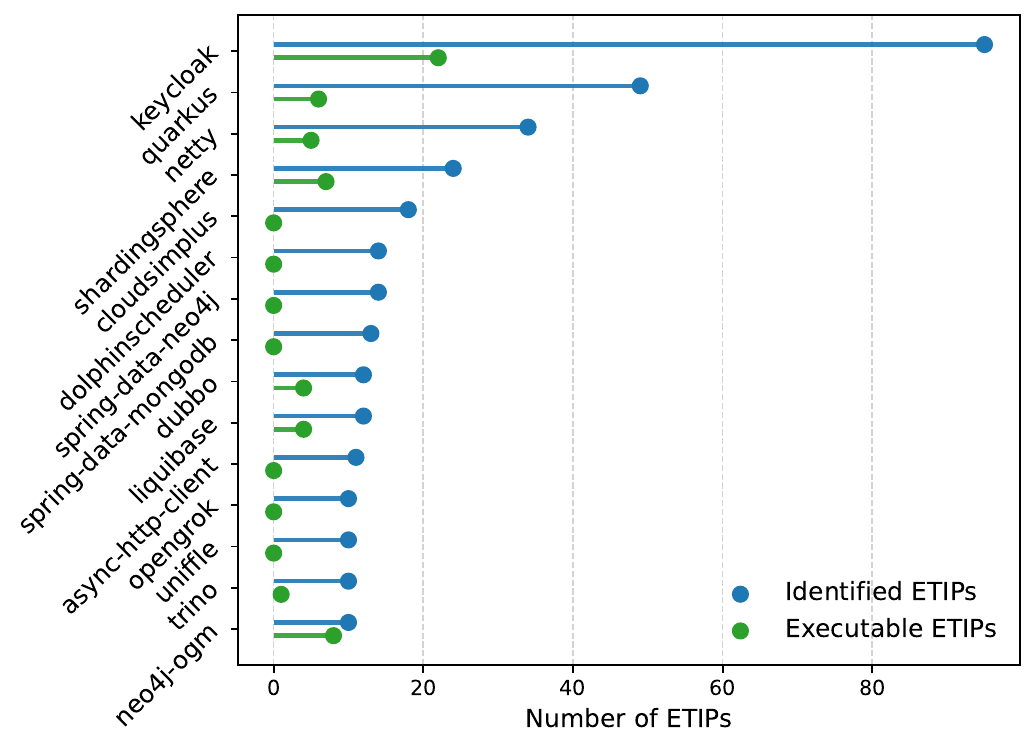}
\caption{The number of identified ETIPs and executable ETIPs from the most frequent repositories in \db.}
\label{fig:repos}
\end{center}
\end{figure}

The identified ETIPs are collected from 174 distinct repositories, while the executable ETIPs come from 40. \autoref{fig:repos} shows the distribution across the top 15 repositories, which cover topics such as distributed systems, data management, cloud-native frameworks, and software infrastructure, demonstrating the topical diversity of \db.

\begin{table}[h]
\centering
\caption{Number of modules and files modified by patches.}
\begin{tabular}{lcccccc}
\toprule
& \multicolumn{3}{c}{\textbf{Identified ETIPs (660)}} & \multicolumn{3}{c}{\textbf{Executable ETIPs (102)}} \\
\cmidrule(lr){2-4} \cmidrule(lr){5-7}
& \textbf{Min} & \textbf{Avg} & \textbf{Max} & \textbf{Min} & \textbf{Avg} & \textbf{Max} \\
\midrule
Modules & 1 & 1.2 & 13 & 1 & 1.1 & 3 \\
Files   & 1 & 2.1 & 20 & 1 & 2.1 & 17 \\
\bottomrule
\end{tabular}
\label{tab:modification-number}
\end{table}

\autoref{tab:modification-number} presents the modification scope of ETIPs in \db. While most patches modify a single module (average close to 1), the benchmark includes patches spanning up to 13 modules and 20 files. Notably, 270 identified ETIPs and 40 executable ETIPs modify multiple files, making \db suitable for evaluating repo-level patch generation tools~\cite{Jimenez2023SWEbenchCL}.

\begin{figure}
\begin{center}
\includegraphics[width=0.7\columnwidth]{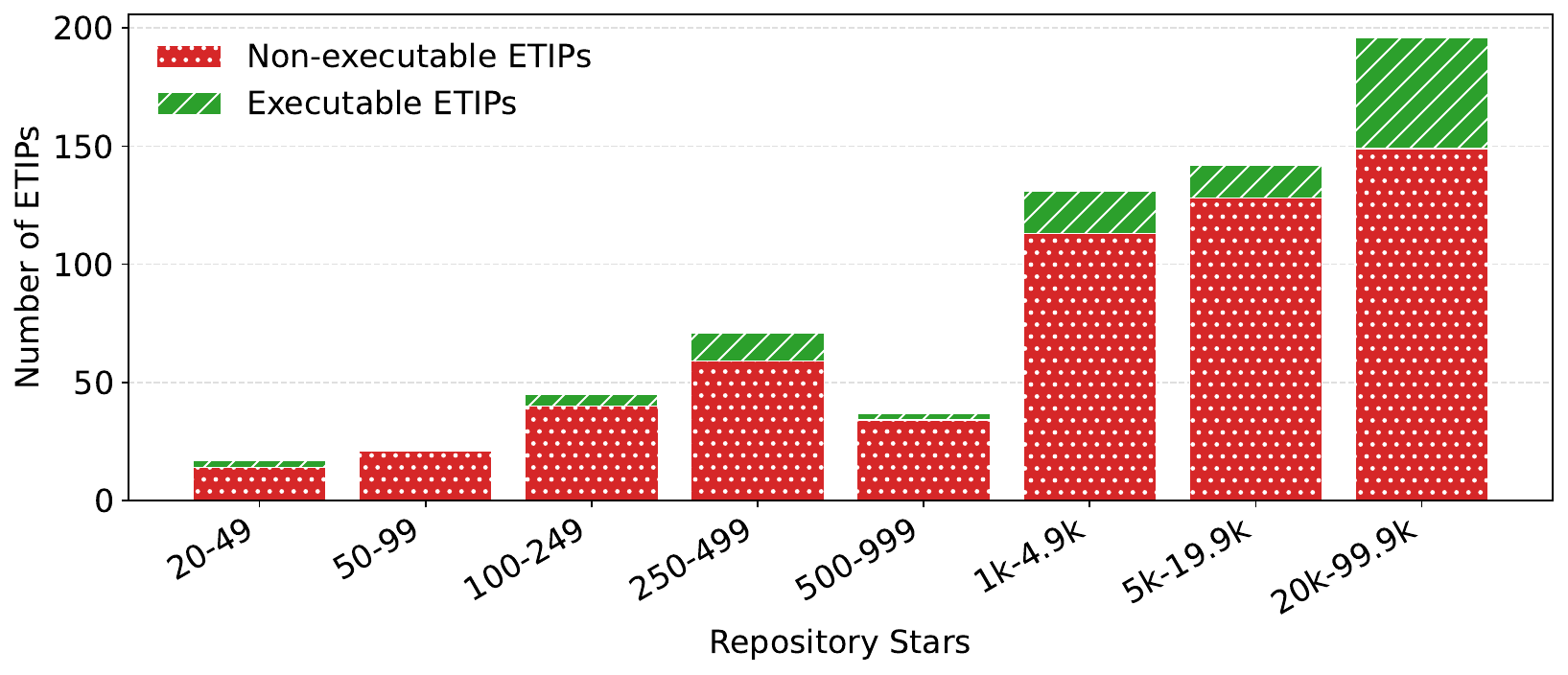}
\caption{The number stars of the repositories from which the ETIPs in \db are collected.}
\label{fig:commits-stars}
\end{center}
\end{figure}

\begin{figure}
\begin{center}
\includegraphics[width=0.7\columnwidth]{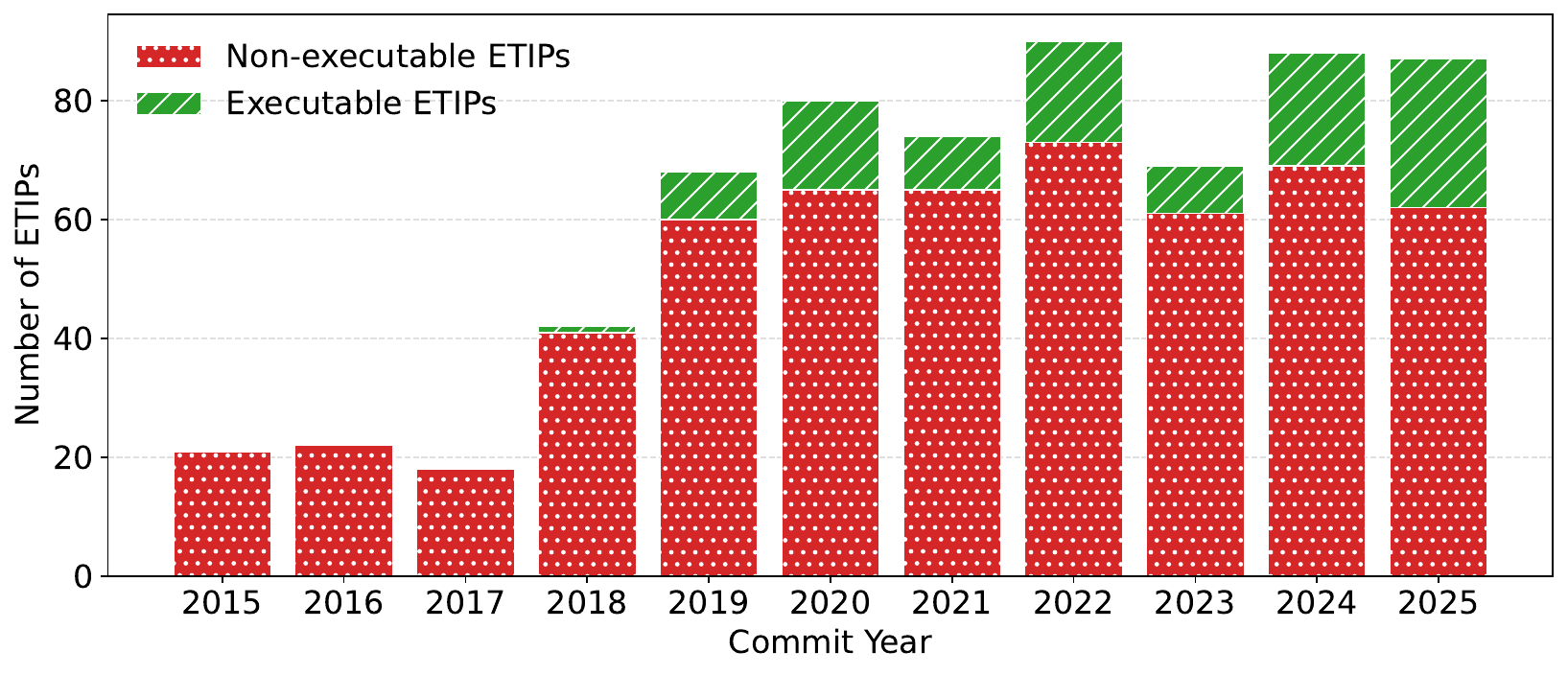}
\caption{The year in which the ETIPs in \db are created.}
\label{fig:commit-year}
\end{center}
\end{figure}

\autoref{fig:commits-stars} shows that most ETIPs come from repositories with at least 10,000 stars, indicating maturity and reliability. \autoref{fig:commit-year} shows a steady distribution of ETIPs from 2019 onward, with fewer from earlier years due to the Java 8 minimum requirement. This temporal diversity ensures the benchmark is not biased toward a particular development period.

Finally, we check if existing JUnit tests serve as ETIP detector tests. With the default configuration ($p\_val=0.1$, $min\_impr=0.05$), only 19 ETIPs have a test class showing statistically significant improvement. \modified{We notice that among these 19 ETIPs, for 6 of them, there is a test class that shows that the original version actually runs statistically significantly faster than the patched version, meaning that the patch is also causing a performance regression under certain conditions. With the conservative configuration ($p\_val=0.05$, $min\_impr=0.1$), only 8 ETIPs have a test class that shows their execution time improvement is statistically significant.} Our takeaway is that open-source Java projects largely lack tests that verify execution time improvements, presenting an opportunity for research on automated ETIP detector test generation.

\begin{mdframed}\noindent
    \textbf{Answer to RQ1: \rqone} \\
    \db includes 660 identified ETIPs from 174 repositories and 102 executable ETIPs from 40 repositories, spanning 11 years of development history and repositories with 20 to 93,448 stars. The benchmark covers diverse project types and includes 270 multi-file ETIPs suitable for evaluating repository-level patch generation tools.
\end{mdframed}

\subsection{Results for RQ2 (using \db in practice)}
\label{sec:rq2-res}

\begin{table*}[t]
  \centering
  \caption{\modified{Results of coding agents on generating execution time improvement patches.
  A patch is deemed a \textbf{Correct Fix} if it is either an
  \textbf{Exact Match} (its AST is identical to the ground truth) or a
  \textbf{Sem.\ Equiv.} (semantically equivalent but syntactically different).
  \textbf{Corr.\ Loc.}~indicates a patch that modifies the correct location but differs semantically,
  \textbf{Wron.\ Loc.}~indicates a patch that modifies an incorrect location, and
  \textbf{Not Gen.}~indicates that the agent failed to produce a patch.}}
  \begin{tabular}{@{}ll r rrr rrr@{}}
    \toprule
    & & & \multicolumn{3}{c}{\textbf{Correct Fix}} & & & \\
    \cmidrule(lr){4-6}
    \textbf{Model} & \textbf{Issue Scope} & \textit{N}
      & \textbf{Exact} & \textbf{Sem.} &
      & \textbf{Corr.} & \textbf{Wron.} & \textbf{Not} \\[-0.3em]
    & & & \textbf{Match} & \textbf{Equiv.} & \textbf{Total}
      & \textbf{Loc.} & \textbf{Loc.} & \textbf{Gen.} \\
    \midrule
    \multirow{5}{*}{GPT-5-mini}
      & \emph{By file scope} \\
      & Single-file    & 54 & 7  & 3 & 10\;(18.5\%) & 19\;(35.2\%) & 7\;(13.0\%)  & 18\;(33.3\%) \\
      & Multi-file     & 37 & 3  & 0 & 3\;(8.1\%)   & 6\;(16.2\%)  & 13\;(35.1\%) & 15\;(40.5\%) \\
      & \emph{By module scope} \\
      & Single-module  & 78 & 8  & 3 & 11\;(14.1\%) & 24\;(30.8\%) & 14\;(17.9\%) & 29\;(37.2\%) \\
      & Multi-module   & 13 & 2  & 0 & 2\;(15.4\%)  & 1\;(7.7\%)   & 6\;(46.2\%)  & 4\;(30.8\%)  \\
      \cmidrule(l){2-9}
      & \textbf{Overall} & \textbf{91} & \textbf{10} & \textbf{3} & \textbf{13\;(14.3\%)} & \textbf{25\;(27.5\%)} & \textbf{20\;(22.0\%)} & \textbf{33\;(36.3\%)} \\
    \midrule
    \multirow{5}{*}{GLM-5.1}
      & \emph{By file scope} \\
      & Single-file    & 54 & 3 & 4 & 7\;(13.0\%) & 12\;(22.2\%) & 3\;(5.6\%) & 32\;
      (59.3\%) \\
      & Multi-file     & 37 & 1 & 1 & 2\;(5.4\%)  & 11\;(29.7\%) & 0\;(0.0\%) & 24\;(64.9\%) \\
      & \emph{By module scope} \\
      & Single-module  & 78 & 4 & 5 & 9\;(11.5\%) & 19\;(24.4\%) & 3\;(3.8\%) & 47\;(60.3\%) \\
      & Multi-module   & 13 & 0 & 0 & 0\;(0.0\%)  & 4\;(30.8\%)  & 0\;(0.0\%) & 9\;(69.2\%)  \\
      \cmidrule(l){2-9}
      & \textbf{Overall} & \textbf{91} & \textbf{4} & \textbf{5} & \textbf{9\;(9.9\%)} & \textbf{23\;(25.3\%)} & \textbf{3\;(3.3\%)} & \textbf{56\;(61.5\%)} \\
    \midrule
    \multirow{5}{*}{Mimo-pro-2.5}
      & \emph{By file scope} \\
      & Single-file    & 54 & 6 & 0 & 6\;(11.1\%) & 11\;(20.4\%) & 3\;(5.6\%) & 34\;(63.0\%) \\
      & Multi-file     & 37 & 1 & 2 & 3\;(8.1\%)  & 12\;(32.4\%) & 3\;(8.1\%) & 19\;(51.4\%) \\
      & \emph{By module scope} \\
      & Single-module  & 78 & 6 & 1 & 7\;(9.0\%)  & 18\;(23.1\%) & 6\;(7.7\%) & 47\;(60.3\%) \\
      & Multi-module   & 13 & 1 & 1 & 2\;(15.4\%) & 5\;(38.5\%)  & 0\;(0.0\%) & 6\;(46.2\%)  \\
      \cmidrule(l){2-9}
      & \textbf{Overall} & \textbf{91} & \textbf{7} & \textbf{2} & \textbf{9\;(9.9\%)} & \textbf{23\;(25.3\%)} & \textbf{6\;(6.6\%)} & \textbf{53\;(58.2\%)} \\
    \bottomrule
  \end{tabular}
  \label{tab:rq2-res}
\end{table*}
 
As explained in \autoref{sec:rq2-protocol}, before running \openhands on \db, we first manually check the executable ETIPs in \db and their corresponding issues. \modified{We only keep the issues that are precisely focused on 
execution time improvement and exclude false positives. The curated set contains 91 issues and their executable ETIPs. The details of this manual analysis are presented in \autoref{sec:rq3-protocol} and \autoref{sec:rq3-res}}. We run \openhands on these 91 cases and prompt it to generate patches to fix the execution time issue in the original version. We then manually assess the generated patches to gain a deep understanding of \openhands performance on \db.

\modified{\autoref{tab:rq2-res} shows the results of fixing the 91 manually verified execution time issues, broken down by the employed LLM (GPT-5-mini, GLM-5.1, and Mimo-pro-2.5) and issue scope (single-file vs.\ multi-file and single-module vs.\ multi-module). Each generated patch is classified as a correct fix (exact match or semantically equivalent), a modification at the correct location but with semantic differences, a modification at the wrong location, or not generated.}

\begin{lstlisting}[float,style=diff, caption={The original code, the ground-truth patch and the \openhands generated patch for issue \#1320 of ``cloud-opensource''. The patches are semantically equivalent, but do not exactly match syntactically.}, label=lst:semantic-equivalence]
%{\color{blue} \textbf{\textbf{Original Version}:}}%
private void readClassFileNames() throws IOException {
  URL jarUrl = jar.toUri().toURL();
  URLClassLoader classLoaderFromJar = new URLClassLoader(new URL[] {jarUrl}, null);
  ClassPath classPath = ClassPath.from(classLoaderFromJar);
  ImmutableSet<ClassInfo> allClasses = classPath.getAllClasses();
  classFileNames = allClasses.stream().map(ClassInfo::getName).collect(toImmutableSet());
}
%\hrule%
%{\color{blue} \textbf{\textbf{Ground-truth Patched Version and Its Diff with OpenHands Patched Version}:}}%
private void readClassFileNames() throws IOException {
  try (JarFile jarFile = new JarFile(jar.toFile())) {
    ImmutableSet.Builder<String> classNames = ImmutableSet.builder();
    Enumeration<JarEntry> entries = jarFile.entries();
    while (entries.hasMoreElements()) {
      JarEntry entry = entries.nextElement();
      String name = entry.getName();
      if (name.endsWith(".class")) {
%\RHilight%+       String className = name.replace('/', '.').substring(0,name.length()-6);
%\GHilight%+       String className = name.substring(0, name.length() - ".class".length()).replace('/', '.');
        classNames.add(className);
      }
    }
  }
  this.classFileNames = classNames.build();
}
\end{lstlisting}

As presented in \autoref{tab:rq2-res}, \openhands generates a correct fix for 14.3\% (13/91) of the issues when GPT-5-mini is employed. This aligns with state-of-the-art studies on fixing performance issues in other programming languages \cite{garg2025perfbench}. This suggests that \db is a reliable benchmark for evaluating advanced patch generation tools on Java execution time issues. The number of correct fixes declines to 9.9\% (9/91) when GLM-5.1 or Mimo-pro-2.5 is used, which shows that GPT-5-mini slightly outperforms the other two.

\modified{For GPT-5-mini, we see that 10 of the correct fixes exactly match the ground-truth, while 3 of them are semantically equivalent to the ground-truth but do not match it.} \autoref{lst:semantic-equivalence} is an example of a semantically equivalent patch generated by \openhands for issue \#1320 of ``cloud-opensource''. This listing shows the original code, the ground-truth patched version and the patched version generated by \openhands. To save space, we show the ground-truth and \openhands patches in a single snippet as a diff. The issue states that in the original version all the classes are loaded from the Jar file and then the class names are taken from the loaded classes. In the patched version, the class names are directly read from the Jar file. The difference between the ground-truth and generated patches appears at lines 19-20, where they build the class name. Both versions read the filename, remove \texttt{.class} from the end of it, and replace \texttt{/} with \texttt{.}, but do this in a different order. Also, the ground-truth uses 6 as a literal, while \openhands computes this literal with \texttt{".class".length()}. This example shows how a generated patch can apply the same improvement as the ground-truth with a different syntactical modification. Consequently, we cannot solely rely on statically checking the changes to the AST. This demonstrates the importance of having more fine-grained and execution-based feedback on generated patches. The \toolname evaluation harness gives such feedback by running generated patch in the docker-based reproducible environment of executable ETIPs.

\begin{figure}[t]
\centering
\includegraphics[width=0.5\linewidth]{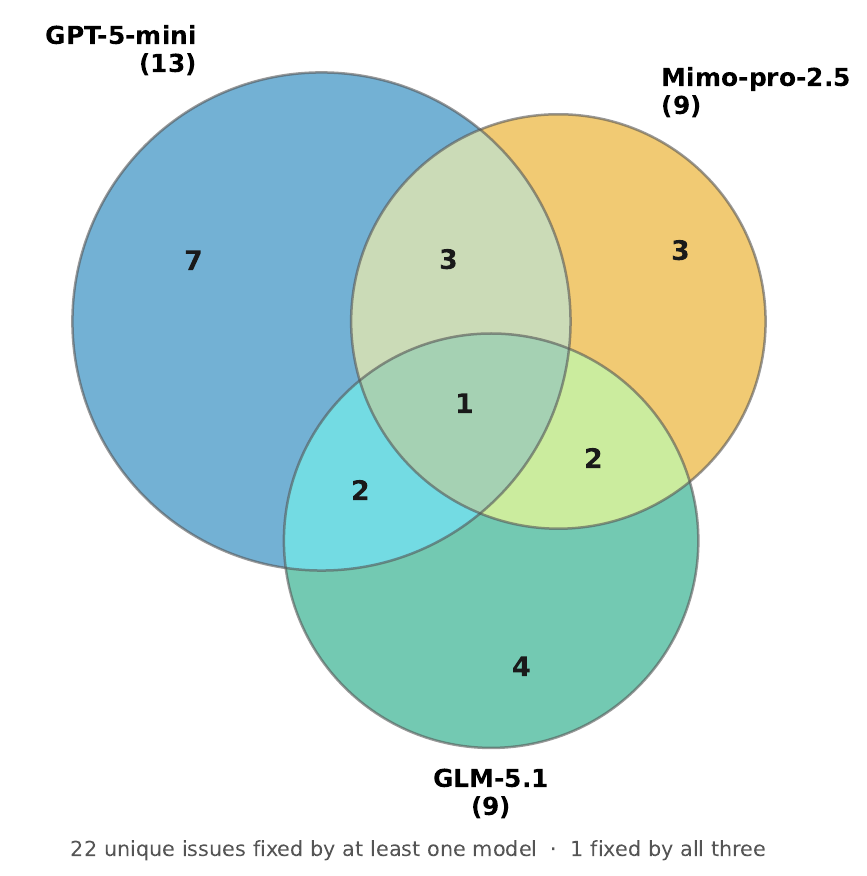}
\caption{Correctly fixed issues by model.}
\label{fig:correct-fixes-venn}
\end{figure}

\modified{\autoref{fig:correct-fixes-venn} shows the relation between correctly fixed issues by each model. Overall, 24\% (22/91) of the issues are fixed by at least one of the models, 7, 3, and 4 of which are fixed only by GPT, Mimo, and GLM, respectively. This shows that each model resolves several unique issues, indicating that none of the models can be fully replaced by the other two. There is also one single issue that is fixed by all three models. This is issue\_\#29319 of the ``Keycloak'' project, which explains that the list that a method returns does not need to be sorted. All models successfully fix the issue by removing the \texttt{sort} call. This suggests that only short and simple performance issues can be fixed by most advanced LLMs, and in general, automatically improving the program execution time requires accurately choosing the agent and its settings.}

\modified{\autoref{tab:rq2-res} shows that for 36.3\% (33/91) of issues, \openhands does not generate a patch, when GPT is employed. For GLM and Mimo, failure rate is even higher at 61.5\% and 58.2\%. Looking at \openhands logs, we find that in most of these cases \openhands fails to generate a reasonable patch because it cannot understand the context of the issue and find the exact location of the fault. This indicates the importance of fault localization for fixing execution time issues.
Using GPT, \openhands fails to generate a correct patch in 27.5\%  of cases, while it localizes the fault and modifies the correct location. This is a promising result that shows the feasibility of improving the effectiveness of \openhands on \db.}

\modified{The comparison between issues with different scopes in \autoref{tab:rq2-res} shows that \openhands can fix issues at all scopes: using GPT, 18.5\% of single-file issues, 8.1\% of multi-file issues, 14.1\% of single-module issues, and 15.4\% of multi-module issues are fixed.} This indicates the ability of \openhands to fix repository level bugs and the usefulness of \db in evaluating repository level bug fixing, which is becoming the focus of patch generation research \cite{Jimenez2023SWEbenchCL}. 
Examining the two multi-module correct fixes, we find that both involve small, repetitive changes across modules (\eg changing a constant from \texttt{RefreshPolicy.WAIT\_UNTIL} to \texttt{RefreshPolicy.IMMEDIATE}), suggesting that complex multi-module performance issues remain an open challenge.

\modified{Finally, we use the harness of \toolname to apply the generated patches, compile the project, and run the existing tests. We find that all patches that are labeled as correct successfully pass the tests, which works as a sanity check confirming that the patch does not change the functionality of the program. We also find that most incorrect patches pass the test, suggesting that they keep the functionality of the program intact too. These patches are incorrect as they fail to recognize the main performance issue and apply the intended patch to fix it.}

\begin{mdframed}\noindent
    \textbf{Answer to RQ2: \rqtwo} \\
    \modified{Employing GPT-5-mini, \openhands correctly fixes 14.3\% (13/91) of the manually verified execution time issues in \db, aligning with results reported for other programming languages \cite{garg2025perfbench}. For both GLM-5.1 and Mimo-pro-2.5, the success rate drops to 9.9\% (9/91), suggesting the superiority of the GPT model on our benchmark. We also see that GPT, GLM, and Mimo fix 7, 4, and 3 unique issues that are not fixed by any other model, respectively, and only one issue is fixed by all models. This indicates that the tasks in \db can only be fixed when an appropriate agent and setting is used. \openhands (using GPT-5-mini) also generates correct fixes for 3 multi-file and 2 multi-module issues, demonstrating the applicability of \db for evaluating repository-level patch generation tools.}
\end{mdframed}

\subsection{Results for RQ3 (\toolname performance)}
\label{sec:rq3-res}

\begin{figure*}
  \centering
  \begin{subfigure}[b]{0.48\textwidth}
    \centering
    \includegraphics[width=\linewidth,height=7cm,keepaspectratio]{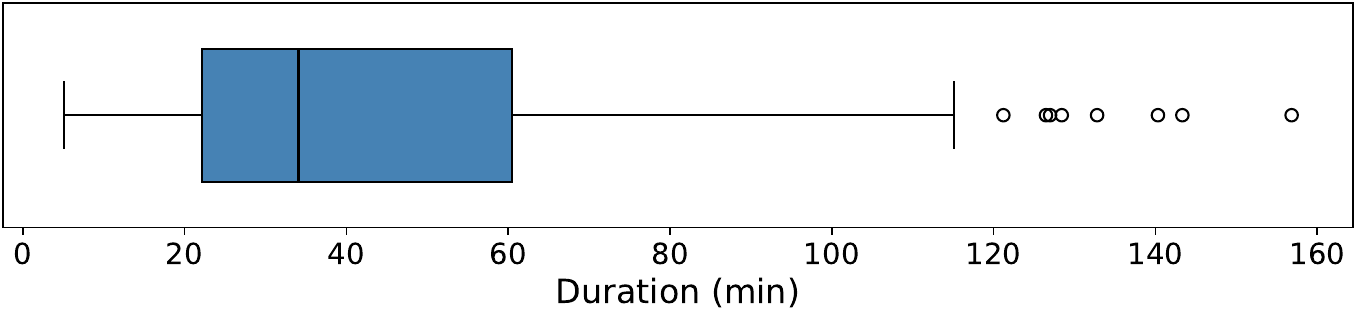}
    \caption{Distribution of time to build the images (min.).}
    \label{fig:commit-analysis-duration}
  \end{subfigure}
\hfill
  \begin{subfigure}[b]{0.48\textwidth}
    \centering
    \includegraphics[width=\linewidth,height=7cm,keepaspectratio]{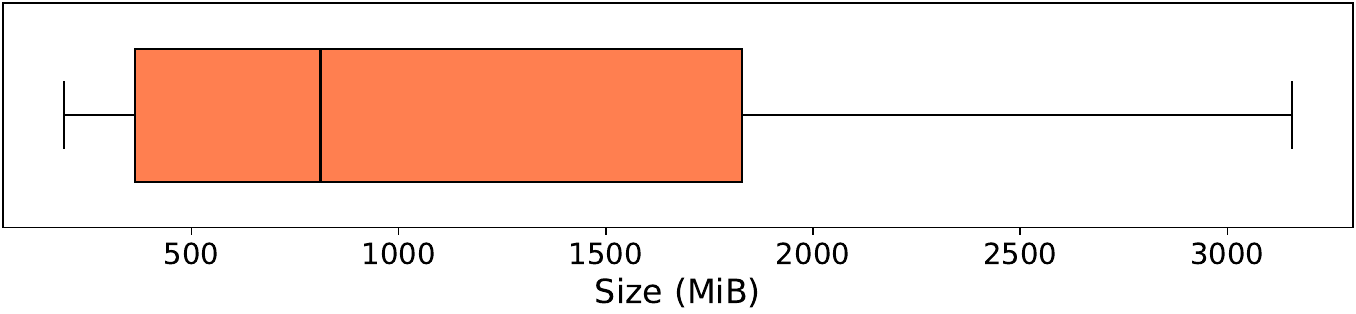}
    \caption{Distribution of the size of images (MiB).}
    \label{fig:ghcr-image-sizes}
  \end{subfigure}
  \caption{Characteristics of Docker images built by \toolname.}
  \label{fig:durations-and-image-sizes}
\end{figure*}

\modified{As explained in \autoref{sec:rq3-protocol}, to assess the precision of \toolname, we manually analyze all 660 identified ETIPs. This analysis consists of a manual inspection by one of the authors, an automated labeling by GPT-5.5, and, if needed, a tie breaking examination by an external expert. Note that the LLM-based filter of \toolname considers all issues corresponding to these ETIPs to be focused on execution time. Our thorough analysis reveals that 518 of these issues are focused on execution time improvement, which also includes 91 of the executable ETIPS as mentioned in \autoref{sec:rq2-res}. This yields a precision of 78\% (518/660) for \toolname, which is aligned with the precision of recent issue classification techniques \cite{aracena2025applying}, making \toolname a reliable tool for collecting ETIPs.}

\begin{figure}
\begin{center}
\fbox{\includegraphics[width=0.8\columnwidth]{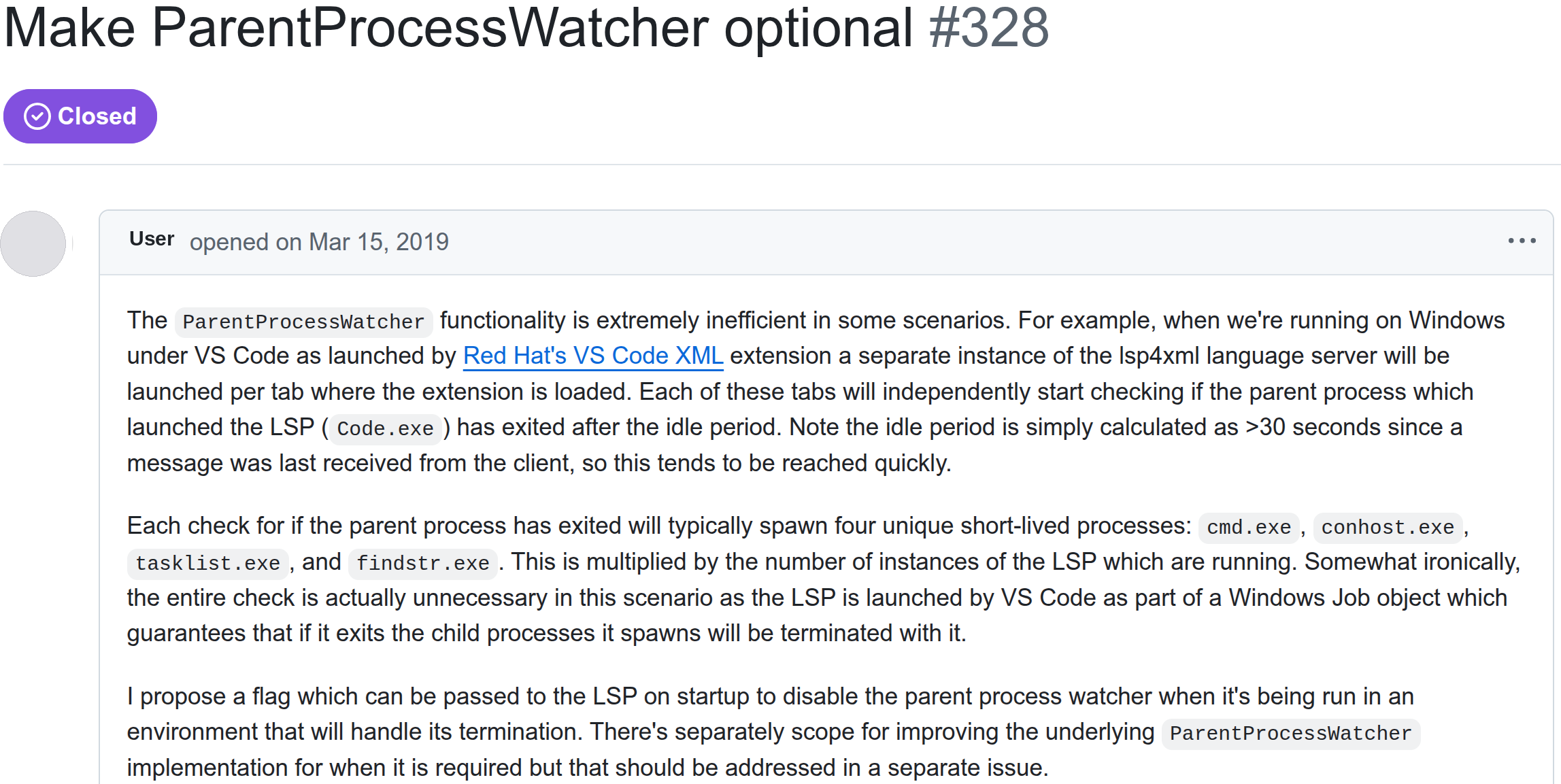}}
\caption{\modified{Issue\_\#328 of the ``lemminx'' project. The issue asks for a new input that makes using \texttt{ParentProcessWatcher} optional and improves the performance. While the issue is focused on execution time improvement, it is an incorrectly identified ETIP, as it requests changes to the program functionality in terms if input/output.}}
\label{fig:precision-feat}
\end{center}
\end{figure}

\modified{In our manual analysis of the identified ETIPs, we find that one of the major sources of false positives are the issues that actually focus on execution time improvement, but also require implementing a new functionality. There are 27 issues of this type, which falls outside our ETIP definition (\autoref{sec:background}). For example, \autoref{fig:precision-feat} shows issue\_\#328 in the ``lemminx'' project\footnote{\url{https://github.com/eclipse-lemminx/lemminx/issues/328}}, which asks for a new functionality to make using \texttt{ParentProcessWatcher} optional, as it is ``extremely inefficient'' in some scenarios. The patch resolving this issue improves the execution time in certain cases, but also changes the input of the program by adding a flag that disables \texttt{ParentProcessWatcher}. We consider cases like this as incorrectly identified ETIPs, as they do not preserve functional equivalence. However, we note that these ETIPs can still be useful for evaluating performance optimizing patch generation tools if those tools are allowed to introduce new features to improve the code.}

\modified{To assess overall accuracy of \toolname, we also randomly sample 100 issues passed to the ``LLM-based Issue Filter'' step. Our analysis shows that 94 of these issues are not focused on execution time improvement and \toolname also correctly labels as such. According to our manual analysis, only six of the issues focus on improving execution time, indicating the scarcity of such issues and the value of the datasets comprising them, like \db. The LLM-based filter of \toolname correctly  detects one of these issues. This experiment shows that the accuracy of \toolname is 95\% (95/100), further indicating the reliability of \toolname.}

Measuring the time dedicated to running the static analysis phase of \toolname on the 1,769,958 target commits, we find that \toolname takes 478 hours to identify the 660 ETIPs. This shows the value of \db for future research on fixing execution time issues, as it provides filtered data that requires almost 20 days of commit inspection. The static analysis of the commits also spends \$5.33 to filter the issues with the help of LLMs, which shows the affordability of our approach.

The dynamic analysis phase of \toolname takes 152 hours to process the 660 identified ETIPs and create 102 executable ETIPs in form of Docker images. This means that processing each identified ETIP takes only 13.8 minutes, which indicates the feasibility of building large benchmarks with \toolname. \autoref{fig:durations-and-image-sizes} shows the characteristics of the Docker images built by \toolname. The median time it takes to process an identified ETIP and build the Docker image is 34 minutes. Also, the median size of built docker images is 790 MB. Again, this shows that \toolname can be used in practice to create Docker images for reproducible execution of ETIPs.

\begin{mdframed}\noindent
    \textbf{Answer to RQ3: \rqthree} \\
    \modified{\toolname achieves 78\% precision in identifying execution time issues, based on manual analysis of all 660 identified ETIPs.} The static analysis phase processes 1.7 million commits in 478 hours at a cost of \$5.33, while the dynamic analysis phase builds Docker images with a median time of 34 minutes and a median size of 790\,MB. These results indicate that \toolname is both reliable and practical for building large scale ETIP benchmarks.
\end{mdframed}

\section{Discussion}
\label{sec:discussion}

\subsection{JMH Tests}
\label{sec:jmh}

As explained in \autoref{sec:dynamic-analysis}, \toolname runs the JUnit tests multiple times and performs statistical testing to see if any test detects a significant execution time improvement. Our experiments show that such JUnit tests are rare (see \autoref{sec:rq1-results}). This leads us to another question:

\begin{quote}
\emph{Is there a framework that open-source Java projects commonly use to demonstrate execution time improvements?}
\end{quote}

The standard and most widely used framework for performance testing in Java is the Java Microbenchmark Harness (JMH)\footnote{\url{https://github.com/openjdk/jmh}}~\cite{yi2025experimental}. We analyze the 91 manually verified executable ETIPs in \db to assess the use of JMH in open-source Java projects. We find that only 15\% (14/91) of the executable ETIPs have JMH benchmarks in their code. This shows that even considering JMH benchmarks does not solve the lack of tests that demonstrate execution time improvements. Therefore, we emphasize the importance of adding more ETIP detector tests in open-source Java projects.

\subsection{Test Generation}
\label{sec:test-generation}

Given the lack of tests that show execution time differences in open-source projects, automated ETIP detector test generation is a valuable task for future research. As explained in \autoref{sec:evaluation-harness}, the evaluation harness of \toolname enables the evaluation of tools that generate ETIP detector tests. \modified{In a preliminary study, we employ \openhands (with GPT-5-mini) to generate tests that show a statistically significant difference between the execution times of original and patched versions for our 91 manually verified executable ETIPs.} We find that \openhands is able to generate an ETIP detector test for only one of the 91 ETIPs (the generated test and the corresponding ETIP are available in our replication package~\cite{repo}). This indicates the difficulty of generating ETIP detector tests and presents a promising opportunity for future studies.

The challenge of generating ETIP detector tests lies in the fact that such tests must not only exercise the modified code paths, but must do so with inputs large or representative enough to reveal measurable performance differences. A test that calls the improved code with a small input may pass on both versions with little timing difference, even if the patch provides orders of magnitude improvement on realistic workloads. This contrasts with functional test generation, where a single well-chosen input can definitively distinguish correct from incorrect behavior. Future work on ETIP detector test generation will likely need to combine code coverage techniques with workload synthesis to produce inputs that stress the performance-critical paths identified by the patch.

\subsection{Threats to Validity}
\label{sec:threats}

\subsubsection{Threat to External Validity}
\label{sec:external-threats}

\toolname collects ETIPs from projects that use Maven Wrapper to build. This means Java projects that use other build tools, such as Gradle, are not considered. Hence, we cannot be certain that our approach generalizes to Java projects that do not use Maven Wrapper. However, as reported in \autoref{sec:rq1-results}, the number of ETIPs identified by \toolname aligns with similar benchmarks for other programming languages~\cite{garg2025perfbench,he2025swe}, which indicates the usefulness of our tool for evaluating advanced patch generation tools.

\subsubsection{Threat to Internal Validity}
\label{sec:internal-threats}

The dynamic analysis and evaluation harness of \toolname run project tests to determine if they detect an execution time improvement. Since the execution environment can be volatile, the results of this step may not be fully robust and reproducible. To address this concern, we run the tests multiple times and use statistical testing to report how confidently we can claim a test detects an improvement. However, in our experiments, we noticed that the order of running tests can affect execution time. This is an issue because by default, Maven does not guarantee a deterministic order for running tests. We resolve this with two techniques. First, we always configure Maven to run the tests in alphabetical order on both versions. Second, over the $num\_exec$ rounds of executing tests on original and patched versions, at odd rounds we first run the patched version and at even rounds we first run the original version. These two techniques minimize the effects of test execution order and environment volatility to maximize confidence in our statistical test results.

\section{Related Work}
\label{sec:related-work}
Recently, many researchers are working on automated repair of various non-functional bugs, including performance issues \cite{garg2025rapgen,ren2025peace,yang2025perfcoder,garg2022deepdev,singhal2024nofuneval,gong2025language}. This line of research heavily depends on having benchmarks for the evaluation of patches generated by repair tools. Consequently, researchers are also working on new performance improvement benchmarks and tools to collect such benchmarks \cite{shetty2025gso,he2025swe,garg2025perfbench}. In this section, we review the literature on performance issue benchmarks, automated fixing of performance issues, and studies on the performance optimization capabilities of AI agents, which are the closest research areas to our work.

\subsection{Performance Issue Benchmarks}
\label{sec:perf-benchmarks}
There are established bug benchmarks, such as Defects4J~\cite{just2014defects4j}, Bugs.jar \cite{saha2018bugs}, IntroClassJava~\cite{durieux2016introclassjava}, QuixBugs~\cite{lin2017quixbugs}, and Bears~\cite{madeiral2019bears}, which have been used for evaluating automated program repair (APR) tools for more than a decade. However, these benchmarks are particularly focused on functional bugs and rarely updated. In a more recent effort, Jimenez \etal~\cite{Jimenez2023SWEbenchCL} have proposed SWE-Bench, which has become one of the most prominent benchmarks for APR evaluation. SWE-Bench provides pairs of GitHub issues and their resolving PRs as well as Docker containers for reproducing the issues and executing generated patches. In contrast to \db, SWE-Bench is focused on Python and functional bugs.

After years of progress in automatically fixing functional bugs, researchers have recently paid more attention to building APR tools for performance bugs \cite{ma2025swe}. Consequently, new benchmarks of performance bugs are also created to support this effort \cite{blot2025comprehensive}. Several benchmarks are collected from code competition submissions, including PIE~\cite{Madaan2023LearningPC} (77K C++ modifications), EffiBench~\cite{huang2024effibench}, EvalPerf~\cite{liu2024evaluating}, and Mercury~\cite{du2024mercury}. While competition submissions are widely used for evaluating the performance of generated code \cite{coignion2024performance}, these programs are usually small and different from real-world projects. To address this limitation, researchers have created more realistic benchmarks from real-world repositories: SWE-Perf~\cite{he2025swe} (140 tasks from 12 Python repositories), SWE-fficiency~\cite{ma2025swe} (performance PRs from nine data-science and HPC repositories), GSO~\cite{shetty2025gso} (102 tasks from 10 codebases in Python, C, and SIMD), PeacExec~\cite{ren2025peace} (146 tasks from 47 Python repositories), and PerfBench~\cite{garg2025perfbench} (81 tasks from .NET repositories with an evaluation harness for test generation). Evaluations on these benchmarks consistently show the difficulty of automatically fixing performance issues, with \openhands achieving success rates of 4.9\% on GSO and 3\% on PerfBench. Automated performance improvement tools have also been tested on more specialized programs, such as GPU kernels~\cite{ouyang2025kernelbench} and compilers~\cite{cummins2023large}.

Compared to these benchmarks, \db is the first benchmark of execution time improvement patches specifically targeting Java programs. While the existing benchmarks primarily focus on Python~\cite{he2025swe,ma2025swe,ren2025peace}, C++~\cite{Madaan2023LearningPC}, .NET~\cite{garg2025perfbench}, or multiple languages~\cite{shetty2025gso}, Java remains underrepresented despite being one of the most widely used programming languages. The most closely related benchmark that focuses on Java is the dataset used by Yi and Leitner~\cite{yi2025experimental}, which contains 65 tasks from four open-source Java projects. However, their dataset does not provide an automated pipeline for collecting new tasks or Docker-based environments for reproducible evaluation. \toolname provides a fully automated pipeline for creating new benchmarks with user-defined filters, Docker-based reproducible environments, and an evaluation harness that supports both patch and test generation evaluation.

\subsection{Fixing Performance Issues}
\label{sec:perf-fix}

The existing benchmarks have enabled many studies on automated fixing of performance issues. Several tools target function-level optimizations: RAPGen~\cite{garg2025rapgen} uses retrieval-augmented prompt generation, DeepDev-PERF~\cite{garg2022deepdev} fine-tunes a transformer model for repairing performance bugs, PerfCoder~\cite{yang2025perfcoder} fine-tunes an LLM on curated real-world optimizations, SBLLM~\cite{gao2025search} combines LLMs with evolutionary search, and EFFI-LEARNER~\cite{huang2024effilearner} uses execution overhead profiles to improve efficiency of LLM-generated code. At the repository level, Ren \etal propose PEACE~\cite{ren2025peace}, a hybrid framework for code efficiency optimization through automatic code editing, achieving a 69.2\% correctness rate on PeacExec.

Researchers have also evaluated the performance of patches generated by agentic frameworks, such as \openhands~\cite{wang2025openhands}, Agentless~\cite{xia2025demystifying}, SWE-Agent~\cite{yang2024swe}, and AutoCodeRover~\cite{zhang2024autocoderover}. Fan \etal~\cite{Fan2025SWEEffiRS} introduce new metrics to compare the performance of programs generated by these frameworks, while Peng \etal~\cite{peng2025agents} present an empirical study comparing agent-authored and human-authored optimizations, finding that the optimization patterns largely overlap.

The work of Yi and Leitner~\cite{yi2025experimental} presents an investigation on LLM-based performance improvements using their dataset of 65 real-world tasks mined from four open-source Java programs. They generate patches using two leading LLMs under four prompt variations and use JMH benchmarks to compare the results with human-authored solutions. Their study demonstrates that human developers outperform LLM fixes by a statistically significant margin.

Our evaluation of \openhands on the 91 manually verified executable ETIPs in \db aligns with the findings reported by these studies. Similarly to previous work, we observe that state-of-the-art agents achieve limited success rates on performance optimization tasks, motivating the need for dedicated benchmarks like \db.

\section{Conclusion}
\label{sec:conclusion}

In this paper, we present \toolname, the first configurable and reusable tool for creating reproducible benchmarks of execution time improvement patches (ETIPs) in real-world Java projects. \toolname combines three phases: 1) static analysis for identifying ETIPs from GitHub repositories, 2) dynamic analysis for building Docker-based reproducible environments, and 3) evaluation harness for quantitative assessment of generated patches and tests.

Using \toolname, we built \db, a benchmark of 660 identified ETIPs and 91 manually verified executable ETIPs collected from 174 open-source Java repositories. \modified{Our manual analysis shows that \toolname identifies ETIPs with a precision of 78\% and an accuracy of 95\%}. Running \openhands on the 91 verified executable ETIPs, we find that it correctly fixes 14.3\% (13/91) of the issues when GPT-5-mini is employed, aligning with results reported by similar studies on other programming languages. We also run \openhands with GLM-5.1 and Mimo-pro-2.5, both of which fix 9.9\% (9/91) of the issues, indicating the superiority of GPT-5-mini on our \db.

Our analysis also reveals that open-source Java projects largely lack tests that demonstrate execution time improvements: only 15\% of verified executable ETIPs contain JMH benchmarks, and existing JUnit tests rarely detect statistically significant execution time differences. Therefore, the evaluation harness of \toolname presents an opportunity for future research on automated fixing of execution time issues and automated generation of ETIP detector tests.

\section*{Acknowledgements}
We used Cursor as a coding assistant during the development of \toolname. We also used Claude to help polish the text and figures of this paper.

\bibliographystyle{ACM-Reference-Format}
\balance
\bibliography{references}

\end{document}